\documentclass[conference]{IEEEtran}

\usepackage{times,graphicx}
\usepackage{balance,url,textcomp,mathtools}
\usepackage[linesnumbered,ruled]{algorithm2e}
\usepackage{color}

\ifodd 1
\newcommand{\com}[1]{\textbf{\color{blue} (COMMENT: #1)}}
\else

\newcommand{\com}[1]{}
\fi

\def\Section {\S}

\newcommand{\squishlist}{
 \begin{list}{$\bullet$}
  { \setlength{\itemsep}{0pt}
     \setlength{\parsep}{3pt}
     \setlength{\topsep}{3pt}
     \setlength{\partopsep}{0pt}
     \setlength{\leftmargin}{1.5em}
     \setlength{\labelwidth}{1em}
     \setlength{\labelsep}{0.5em} } }

\newcommand{\squishlisttwo}{
 \begin{list}{$\bullet$}
  { \setlength{\itemsep}{0pt}
     \setlength{\parsep}{0pt}
    \setlength{\topsep}{0pt}
    \setlength{\partopsep}{0pt}
    \setlength{\leftmargin}{2em}
    \setlength{\labelwidth}{1.5em}
    \setlength{\labelsep}{0.5em} } }

\newcommand{\squishend}{
  \end{list}}

\begin{document}

\title{Hedge Your Bets: Optimizing Long-term Cloud Costs by Mixing VM Purchasing Options}
\author{Pradeep Ambati, Noman Bashir, David Irwin, Mohammad Hajiesmaili, and Prashant Shenoy\\
University of Massachusetts Amherst}
\date{}
\maketitle

\begin{abstract}
Cloud platforms offer the same VMs under many purchasing options  that specify different costs and time commitments, such as on-demand, reserved, sustained-use, scheduled reserve, transient, and spot block. In general, the stronger the commitment, i.e., longer and less flexible, the lower the price.  However, longer and less flexible time commitments can increase cloud costs for users if future workloads cannot utilize the VMs they committed to buying. Large cloud customers often find it challenging to choose the right mix of purchasing options to reduce their long-term costs, while retaining the ability to adjust capacity up and down in response to workload variations.  

To address the problem, we design policies to optimize long-term cloud costs by selecting a mix of VM purchasing options based on short- and long-term expectations of workload utilization. We consider a batch trace spanning 4 years from a large shared cluster for a major state University system that includes 14k cores and 60 million job submissions, and evaluate how these jobs could be judiciously executed using cloud servers using our approach.  Our results show that our policies incur a cost within 41\% of an optimistic  optimal offline approach, and 50\% less than solely using on-demand VMs.


\end{abstract}

  
\section{Introduction}
\label{sec:introduction}
With the advent of cloud computing, large institutions that have traditionally operated large private compute clusters for their general computing needs have begun to migrate to public Infrastructure-as-a-Service (IaaS) cloud platforms, which rent VMs to users for a price per unit time, due to their low cost and high accessibility.  Migrating a large-scale private computing infrastructure to a public cloud  is a complex task. For example, the major hyperscale public cloud providers---Amazon, Google, and Microsoft---now offer dozens of VM types with different CPU, memory, I/O, and network characteristics at different prices.  Thus, selecting the ``right'' type of VM---that yields the desired performance at the lowest cost---for a particular workload can be challenging.  In addition, a significant fraction of a private cluster's cost is due to upfront capital expenses, e.g., server hardware, building space, supporting IT equipment, etc., that are fixed, while a cloud-based cluster's cost is largely operational expenses that are  dependent on expectations of the future workload.  If cloud resources are provisioned judiciously---by choosing the optimal mix of VM options---when migrating large workloads to the cloud, cloud costs can be often lower than operating a private cluster.  However, the large number of cloud VM configurations, the uncertainty in future workload characteristics, and the complexity of decision making when operating a large cloud  cluster imply there is no guarantee that the potential cost savings from migrating to the cloud will actually be realized.

In addition to offering dozens of VM types, cloud platforms also offer these VMs under a variety of {\em purchasing options} that specify different prices and time commitments.  For example, the \emph{on-demand} option is the most common, enabling users to request and release VMs at any time and only be charged for the time they hold them.  In contrast, the \emph{reserved} option requires users to commit to buying 1 or 3 years of VM time in advance, but at a significant discount compared to holding an on-demand VM for 1 or 3 years. Of course, if reserved VMs are utilized less than their discount, then they will incur a higher cost than utilizing the equivalent on-demand VMs, since users may release on-demand VMs when not in use. 

Cloud providers offer numerous other VM purchasing options that specify different time commitments and costs for the same types of VMs, including sustained-use, scheduled reserved, spot/preemptible, and spot block.  The specific set of purchasing options, as well as their names, prices, and some details, differ across cloud providers, which we discuss in \Section\ref{sec:background}. In general, though, the stronger a user's time commitment, i.e., longer and less flexible, the lower the VM's price. However, as illustrated above, despite a lower price, longer and less flexible commitments can increase cloud costs if users' future workload cannot utilize the VMs they committed to buying. Thus, optimizing for long-term cloud costs not only requires selecting the right VM types based on a workload's resource usage, but also selecting the right mix of purchasing options based on future workload expectations.  While the former problem 
of choosing cloud VM types has been the subject of much research~\cite{cherrypick,bestvm2,ernest,bestvm},  optimizing the purchasing options for the chosen VMs has not seen as much attention.  This problem requires cloud users to balance two competing tradeoffs---making the longest possible commitments for the provisioned VMs to extract the greatest savings, while retaining some ability to make short-term increases or decreases to provisioned VMs to respond to changing workloads. 

This paper focuses on optimizing the mix of purchasing options for cloud VMs.  We develop policies to estimate and optimize long-term cloud costs on the major cloud platforms by selecting the mix of available purchasing options based on short- and long-term expectations of workload utilization.  In general, these policies normalize the cost of each VM purchasing option subject to its expected utilization based on historical workload data, and then selects the option for each unit of demand that yields the lowest cost. Further, the uncertainty in future workload demand also needs to be considered when selecting a purchasing option.  As we show, the normalization and uncertainty issues differ for each purchasing option, making the problem non-trivial. For example, for long-term options, such as scheduled reserved, we must consider thousands of options with different costs, e.g., based on when VMs are scheduled to run, respectively, while for a short-term option, such as spot block, the uncertainty of job runtime predictions needs to be considered. 

Our hypothesis is that mixing VM purchasing options enables cloud users to ``hedge their bets'' and minimize their long-term cloud costs relative to exclusively using a \emph{single} purchasing option, such as all on-demand VMs or all reserved VMs. We evaluate this hypothesis and our policies using 4 years of batch traces consisting of 60 million job submissions from a 14k-core shared batch cluster for a major U.S. state university system. Our evaluation considers a scenario where this large workload is migrated to, and operated in, the cloud, and compares realistic online policies to an optimistic  optimal offline variant, which assumes perfect future knowledge of workload demand and a perfectly elastic workload that can be freely split across VMs.  In doing so, our work answers a number of interesting questions, listed below.

\squishlist

\item What is the cost savings from mixing VM purchasing options relative to exclusively using on-demand VMs or reserved VMs? 


\item How do the policies needed to optimize for provider-specific VM purchasing options  differ across cloud providers? 

\item Can users get more cost savings from cloud providers that offer more purchasing options, such as Amazon, tailored to specific types of workloads versus those that offer fewer, such as Microsoft?

\squishend

Our results show that our policy, which mixes VM purchasing options, incur a cost within 41\% of an optimistic optimal offline approach, and are 50\% and 79\% less than solely using on-demand VMs or reserved VMs, respectively. Our work also shows that, while cloud platforms offer numerous complex VM purchasing options, near-optimal procurement strategies for general batch workloads can be simple in practice, as ours largely uses a mix of transient, on-demand, and reserved options available from all of the major cloud providers.

\section{Cloud Provider Overview}
\label{sec:background}
We first provide details on the VM purchasing options offered by the major public cloud providers: Microsoft, Google, and Amazon.  We then discuss differences between the set of options offered by each cloud provider. 

\begin{table*}[t!]
\small
\begin{center}
\begin{tabular}{|l|c|c|c|c|c|}
\hline
\textbf{Purchasing Option} & \textbf{Relative Cost (\%)} & \textbf{Time Commitment} & \textbf{Revocable} & \textbf{Guaranteed} & \textbf{Cloud Providers}
\\
\hline
\hline
{\it On-demand} & 100\% & None & No & No & All\\ \hline
{\it Reserved} & 60\% & 8760hrs (1yr) & No & Yes & All\\ \hline
{\it Reserved} & 40\% & 26,280hrs (3yrs) & No & Yes & All\\ \hline
{\it Transient} & 20-40\% & None & Yes & No & All\\ \hline
{\it Sustained-Use}  & 70-100\% & None & No & No & Google\\ \hline
{\it Customized} & 105\% & - & - & - & Google\\ \hline
{\it Spot Block}  & 55-70\% & None & After 1-6hrs & No & Amazon\\ \hline
{\it Scheduled Reserve} & 90-95\% & 1200-8760hrs (1yr) & No & Yes & Amazon \\ \hline
\end{tabular}
\end{center}
\caption{\emph{Overview of the primary VM purchasing options across the major cloud providers}}
\vspace{-0.5cm}
\label{table:options}
\end{table*}

\subsection{Cloud VM Purchasing Options}

As noted earlier, our work focuses specifically on VM purchasing options that relate to time commitments and flexibility, and not \emph{VM types} or \emph{capacity reservations}.  For example, we do not consider dedicated hosts or VMs, which reduce virtualization overhead and interference, or burstable VMs, which enable VMs to periodically use additional resources.  Since these options only differ in the resources they offer, users can treat them as a different resource type.  The only exception to this, as we discuss below, is customized VMs. We also do not consider capacity reservations, which enable users to pay to ensure that their future requests for on-demand VMs are not rejected due to a lack of capacity.  Currently, these capacity reservations incur the same cost as the on-demand option, so users may just as well purchase and hold on-demand VMs.  We also do not consider some of the small differences between similar purchasing options across cloud providers, although we note some of these differences below. These small differences generally do not affect our policies in \Section\ref{sec:design} or the magnitude of our results in \Section\ref{sec:evaluation}. 
 
Table~\ref{table:options} lists the different VM purchasing options that we consider and their primary attributes.   As can be seen, the {\em same} cloud VM can be procured under a number of different  purchasing options. The relative cost represents the percentage cost relative to the on-demand cost per unit time for the equivalent VM type, and is \emph{not} the percentage discount.  Thus, 60\%  represents 60\% of the on-demand cost, which corresponds to a 40\% discount.  The time commitment is the amount of time the user must commit to buying.  We discuss the other attributes below. 

\noindent {\bf On-demand}. The on-demand option is the most common one offered by all cloud providers, and typically the default option for users.   As a result, we represent the cost of the other options relative to the on-demand option. An on-demand VM incurs a cost per unit time from the time the cloud platform allocates it to the user until the time the user terminates it.  The per unit time cost is now billed at fine-grained resolutions, e.g., either per-second or per-minute, rather than hourly.  Cloud platforms do not generally revoke on-demand VMs, but they are not guaranteed to be available when requested.  That is, cloud platforms may reject users' request for on-demand VMs if they run out of data center capacity.  However, the frequency of out-of-capacity rejections is not publicly known. 

\noindent {\bf Reserved}.  The reserved option enables users to commit to buying a VM for 1 or 3 year period in return for a discount compared to procuring an on-demand VM over the same period.  All cloud providers offer 1 and 3 year reserved options, which are designed for cheaply satisfying a user's expected base load---the minimum level of demand---over the reservation's term. While reserved VMs are not revocable,  they do generally guarantee the user capacity on request. That is, if a user ever terminates a reserved VM, when they request the VM later (within the reservation's term), unlike with on-demand VMs, the cloud platform guarantees to have the capacity to satisfy that request.  Note that Google's reserved option equivalent requires combining a 1 or 3 year committed-use discount with a separate capacity reservation.  As mentioned above, we do not consider purchasing capacity reservations (which Amazon also offers) independent of VMs.

The reserved option is essentially a volume discount, where the actual discount is based on the time commitment as well as other options, such as the amount of upfront payment, whether reserved VMs can be ``converted'' to other VMs of a different type (but the same resources), and whether the reserved VMs can switched between different data centers within the same geographical region.  The costs in Table~\ref{table:options}---60\% and 40\% of the on-demand price for 1 and 3 year terms---are typical discounts for standard options, e.g., payment in full for non-convertible VMs tied to one availability zone.  As expected, the longer the time commitment, the higher the discount.  

\noindent {\bf Transient}.   All cloud providers  offer their suplus capacity in the form of transient VMs~\cite{yank-ieee} but under different names and slightly different terms. Transient VMs are the cheapest purchasing option, costing 20-40\% of the on-demand cost, and come with no time commitment. However, since transient VM resources represent spare capacity, cloud platform's may revoke them at any time to satisfy higher-priority requests for on-demand and reserved VMs.   Given their low cost and priority, transient VMs are not guaranteed, and requests for such VMs are likely rejected due to fluctuating surplus capacity more frequently than on-demand VMs (although the rejection rates are not publicly known).  Transient VMs are generally designed for cheaply satisfying batch jobs that run in the background, and can tolerate delays due to unexpected revocations. 

As with reserved, there are some small differences between Microsoft, Google, and Amazon's transient offering. Microsoft and Google both offer transient VMs, called low-priority batch VMs and preemptible VMs, respectively, for a fixed cost per unit time.  In contrast, Amazon offers spot VMs, which have a variable cost per unit time.  In the past, spot VMs were revoked only when their dynamic spot price, which was often volatile, exceeded a user's bid price.  However, Amazon recently changed spot VMs to remove user bids (and thus decouple revocations from prices), and  reduced price volatility. While spot prices technically remain variable, they are now largely stable.  As a result, spot VMs are now similar to low-priority batch and preemptible VMs.  The key difference in the transient offerings now is that spot VMs (Amazon) and low-priority batch VMs (Microsoft) have no maximum lifetime, while preemptible VMs (Google) have a maximum lifetime of 24 hours after which they are always revoked. 

\noindent {\bf Sustained-Use}.  Google offers a sustained-use discount that automatically applies to on-demand VMs of any type that are run for some fraction of a month-long billing period.  The discount applies separately to each core, i.e., vCPU, and gigabyte (GB) of memory regardless of type, since Google separately charges for each core and GB of memory. Thus, VM types simply incur a cost based on their pre-defined number of cores and memory allotment. The discount starts once each core or GB of memory is used for 25\% of the month, and increases the longer they are used with the maximum discount being 30\% off the on-demand cost for the entire month.  Specifically, for the first 25\% of the month users pay 100\% of the on-demand cost, next for 25-50\% they pay 80\%, then for 50-75\% they pay 60\%, and finally for 75-100\% they pay 40\%.  The overall cost for an entire month of sustained use comes to 70\% of on-demand (i.e., 30\% discount).
The sustained-use discount applies regardless of when a core or GB of memory is used during a month, or whether it is part of a pre-defined VM type. Sustained-use VMs offer a middle option  between  on-demand and reserved VMs, since they cost slightly more than the reserved option but less than on-demand, and also come with no time commitment.

\noindent {\bf Customized}. Google also offers a customized VM option, which enables users to purchase a VM with a custom cost based on a configurable number of cores and memory.  Customized VMs can be used in conjunction with any of the purchasing options above, including the sustained-use discount.  Customized VMs have the potential for significant cost savings by better matching job resource requirements to VM resources, thereby reducing wasted resources.  However, this savings comes at an increased cost, which is currently 105\% the normalized cost per core and GB-memory of an on-demand VM with pre-defined cores and memory allotment.

\noindent {\bf Spot Block}. Amazon offers spot block VMs that have a short pre-defined lifetime of 1, 2, 3, 4, 5, or 6 hours.  Spot block VMs are always revoked after their pre-defined lifetime (but typically not before), although users can terminate them early and only pay for the time they held them.  Thus, spot block VMs have a maximum lifetime, but no time commitment.  Spot block VMs cost 50-70\% of the on-demand cost with higher discounts for shorter lifetimes.  Spot block VMs are a form of {\em short-term reservation} that ensures the cloud platform is able to reclaim resources in the near future.  Their average discount is less than spot, and near that of reserved.  However, spot block VMs do not require a long time commitment, and are designed for short tasks ($<$6 hours) that either have a deadline or cannot gracefully handle revocations, which makes them unsuitable for transient/spot VMs. 

\noindent {\bf Scheduled Reserved}.  Amazon also offers a scheduled reserve option 
 designed for workloads that do not run continuously but do run on a regular schedule, such as  nightly batch jobs  or financial simulations that run after the stock market closes each weekday afternoon. Scheduled reserved VMs enable users to define repeating daily, weekly, or monthly reservation schedules at hourly resolutions.  For example, users could define a daily schedule that reserves a VM from 9pm-12am each day.  As with the reserved option, scheduled reserved capacity is guaranteed and not revocable.  However, the discount is much smaller, only 10\% during off-peak weekend hours and 5\% during peak weekday hours. Scheduled reserved are only offered for a 1 year term, and require users to purchase a schedule with a minimum of 1200 hours over the year. This option is also currently available in only 3 of the larger regions (Northern Virginia, Oregon, and Ireland).  

\subsection{Differences between Cloud Providers}

As Table~\ref{table:options} shows, the major cloud providers offer slightly different VM purchasing options.  Microsoft offers the simplest set with on-demand, 1- and 3-year reserved, and transient options. Google then adds the sustained-use discount for on-demand VMs, along with the ability to configure customized VMs with any purchasing option.  In contrast, Amazon adds the spot block and scheduled reserved options.  Note that Table~\ref{table:options}'s relative cost is an estimate across all the cloud providers.   In general, Microsoft quotes similar prices (in the same way) as Amazon for the same purchasing options and VM types, while Google quotes prices slightly differently.  However, the discounts offered for the purchasing options (even for comparable VMs) are not directly comparable across cloud providers, since aspects of their infrastructure, such as the network and I/O bandwidth, and the resources may differ.  

Our goal is not to analyze which cloud provider offers the lowest absolute cost for each option (for a given workload) based on today's prices, as these prices and each platform's infrastructure can and do change frequently. Rather, our goal is to understand how to choose an appropriate mix of these purchasing options to optimize long-term cloud costs, and to understand their impact on a large-scale batch workload.  Thus, our evaluation in \Section\ref{sec:evaluation} uses the same prices, discounts, and VM types across all the cloud providers based on the estimates in Table~\ref{table:options}.  Finally, our work only considers VM rental costs, and not the additional costs related to network I/O, storage capacity, or the use of other cloud services.

\section{Long-term Cost Optimization}
\label{sec:design}
Given a set of cloud VM types (of different fixed sizes) offered under the purchasing options in \Section\ref{sec:background}, our problem is to select the resources and purchasing options that minimize the long-term cost based on both short- and long-term expectations of workload utilization.   We assume our workload is composed of batch job submissions from users that include the requested number of cores and memory.

To simplify the problem, we first consider an optimistic  optimal offline  approach which assumes perfect knowledge of the future workload as well as the ability to allocate fractional demand to fractional resources where possible. That is, even though our workload's demand is composed of discrete jobs, we assume that discrete jobs can be sub-divided across resources, and purchasing options. Similarly, even though cloud VMs are mostly composed of discrete resource bundles (types), we assume resources (cores and memory) can be purchased separately and bundled together in any quantity.  That is, we assume  resources are allocated in the form of customized VMs, as currently offered by Google, and that the purchasing option price applies separately to cores and memory (as with Google).  As a result, we discuss a workload's utilization in terms of generic units of resource demand, which, in the fractional case, can be either be the number of cores or gigabytes of memory.    We do not separately consider customized versus non-customized options in the offline case. Solving this offline case provides an {\em optimistic} optimal upper bound on the realizable cost savings in practice. We then present our  online approach, which removes the assumptions above  of perfect future knowledge and fractional demand. Our online policy is similar to the offline policy, but substitutes imperfect predictions of short- and long-term demand (based on historical data) for perfect knowledge and considers the availability of limited VM types. 

\subsection{Optimistic Optimal Offline Approach}
\label{sec:offline}

We model the workload trace in terms of the aggregate resource demand per unit time from all active jobs within that time unit; the aggregate resource demand is defined to be the total  cores and memory requested by all active jobs within a time unit. Thus, the workload can be viewed as a time-varying function of  resource  demand. The intuition for our optimal offline  approach is as follows: for each unit of resource demand, e.g., cores and memory requested by jobs, we compute the cost of the necessary resources under each purchasing option to satisfy that unit of demand normalized by its utilization over the length of the commitment.  For example, for one unit of resource demand and a 1-year reservation, we normalize the cost based on the utilization over a year. Thus, if the reservation's cost is 60\% of the on-demand cost, but the utilization over the year is only 60\%, then its normalized cost is the same as the normalized on-demand cost.  Given the normalized costs for various options, we select the cheapest option for each unit of resource demand until the demand across time slots is satisfied.  As discussed below, we apply this approach separately to cores and memory. 

Our general strategy applies directly when considering the relative cost of on-demand, 1- and 3-year reserved, scheduled reserved, and the sustained-use options.  In these cases, we can directly compute a optimistic optimal normalized cost for the resources to satisfy each unit of demand under our assumption of a fractional supply and demand.  However, the normalized cost of the transient and spot block options is \emph{directly} a function of each job's length, which prevents us from directly computing it under the assumption of a fractional resource demand.   For example, for the transient option, the longer the job, the more revocations it will experience and the greater its normalized cost.  Similarly, a job that runs for 3 hours on a 3-hour spot block resources has a higher normalized cost than a job that runs for 1 hour on a 1-hour spot block resource.  Thus, when computing the normalized cost for these options, we \emph{must} consider job length. As a result, in the offline case, we first assign a normalized cost for using the transient and spot block for each job, as discussed below, before considering the other purchasing options.

\noindent {\bf Transient}. As prior work has discussed~\cite{transient-guarantees,spoton}, the normalized cost of using transient VMs is a function of not only their relative cost per unit time and the job's length, but also the revocation rate and the use of fault-tolerance mechanisms to mitigate the impact of revocations.  In general, the longer a job, the more likely it is to experience a revocation. However, precise revocation rates (and their distribution) are not publicly known, and likely differ across providers. For example, Google always revokes preemptible VMs after 24 hours, while historical data suggests Amazon's mean-time-to-revocation for spot VMs may be closer to 48 hours~\cite{flint}. The use of fault-tolerance mechanisms can mitigate the impact of revocations~\cite{transient-guarantees,spoton,flint,exosphere,tr-spark}. For example, if a job employs periodic checkpointing (for some period), on each revocation, it can restart from the latest checkpoint rather than from the beginning.  Of course, employing such mechanisms incurs some overhead that degrades performance and increases the normalized cost.  Prior work has extensively studied these overheads and the normalized cost for transient VMs for different types of jobs, durations, revocation rates, and fault-tolerance mechanisms~\cite{transient-guarantees,spoton,flint,exosphere,tr-spark}. 

Our work can directly apply the prior work above to compute the normalized cost of using transient VMs based on a job's duration and resource usage, as well as the optimal checkpointing frequency based on the revocation rate.  However, we note that such systems-level checkpointing is still not commonly used by large production batch systems for a variety of reasons. For example, while Linux containers include a checkpoint/restart mechanism, it remains under active development, and does not apply to parallel jobs.  Thus, our analysis assumes a more basic use of transient VMs that assumes no checkpointing by restarting a job after each revocation.  

In particular, to ensure a job assigned to a transient VM completes, once it has experienced a revocation, we just restart it on an on-demand VM.  Under this simple model, we compute the expected cost $E[C(T)]$ to execute a job of length $T$ using the transient option as below. We assume that $p_{\texttt{transient}}$ and $p_\texttt{ondemand}$ are the relative transient and on-demand prices, $R(T)$ is the probability the job will be revoked before it completes at time $T$, and $E_\texttt{revoke}[T] < T$ is the expected time a revoked job runs.  

\vspace{-0.3cm}
\begin{equation}
\begin{split}
E[C(T)] = (1-R(T))(p_\texttt{transient} \times  T) + \\
R(T) (p_\texttt{transient}  \times E_\texttt{revoke}[T] + p_\texttt{ondemand} \times T) 
\end{split}
\label{equation:transient}
\end{equation}
The first term represents the total cost to run the job if it is not revoked, while the second term represents the total cost to run the job if it is revoked. 
The normalized cost per unit time is then the expected cost to execute the job $E[C(T)]$ divided by the expected running time, which is $(1-R(T)) \times T + R(T) \times (E_\texttt{revoke}[T] + T)$. Of course, our approach is dependent on the revocation characteristics.  For example, assume a revocation always occurs within 24 hours (as with Google) and that it is equally likely at any point within 24 hours, and that the transient option costs 30\% of on-demand per unit time. In this case, if a job's running time $T$=$18$ hours, then its probability of revocation $R(18)$=$0.75$ and its expected running time if it is revoked $E_\texttt{revoke}[T]$=$9$ hours. As a result, the expected cost $E[C(T)]$=$0.25(0.3)(18) + 0.75((0.3)(9) + (1)(18))$=$16.875$, and the expected running time is $0.25(18) + 0.75(9+18)$=$24.75$ hours, which yields a normalized cost per unit time that is 68\% of the on-demand price. Thus, in this case, the discount relative to using on-demand is only 32\% rather than 70\%.  

Clearly, a shorter job yields a lower normalized cost. For example, a 12 hour job has a normalized cost of 58\% of on-demand. With Google, under our restart model, it never makes sense to run a job with a duration greater than 24 hours, while it may make sense for the other providers, depending on the job's length and revocation characteristics.  For the offline case, we assume we know each job's running time, as well as the revocation characteristics, and can directly compute the normalized cost per job per unit time.  The transient option tends to be the cheapest option for shorter jobs, e.g., a few hours, but can be competitive with other purchasing option for longer jobs.  In our illustrative example, the 12 hour job above has a lower normalized cost than the fully utilized 3-year reserved option (which is 40\% the on-demand price), but the 18 hour job has a higher normalized cost. The transient option is not good for extremely long jobs, e.g., many days or weeks, as the higher revocation probability increases their expected running time and normalized cost.

\noindent {\bf Spot Block}. Spot blocks can be purchased in 1-, 2-, 3-, 4-, 5-, or 6-hour increments with a higher discount applied to shorter increments, such that a 1-hour block is 55\% the on-demand price with each additional hour increasing the price by 3\% resulting in a 6-hour increment that is 70\% the on-demand price.  In the offline case, since we know the duration of each job, we simply map jobs to the smallest spot block increment that is greater than their running time, and compute the corresponding normalized cost per unit of time based on the increment's discount. Jobs longer than 6 hours cannot be run using the spot block option.  As we discuss, for the online case, we must predict each job's running time based on historical data to compute this normalized cost.  This option tends to be slightly more expensive than the transient option, since it is only applicable to short jobs less than 6 hours (where the transient option does well).  As a result, the primary reason to use this option over the transient one is largely based on job requirements.  For example, a job may not be capable of automatically detecting revocations and restarting a job; the job may not be idempotent and thus cannot be re-run after a revocation; or the job may have a fixed deadline that is not amenable to the probabilistic nature of the transient option.

\noindent {\bf On-demand.} Computing the normalized cost for on-demand VMs is straightforward: we simply assign the on-demand cost to each unit of resource demand. 

\noindent {\bf Sustained-Use}.  The sustained-use discount for an on-demand server applies regardless of when a unit of resource demand is used within a month. Thus, to compute it, we need only compute the average resource demand over each month-long billing period.  The full discount applies to the floor of this average, while a partial discount applies to the remainder of the average.   Note that the sustained-use discount applies to the on-demand cost, so, when applied, it always results in a normalized cost equal to or less than the on-demand cost. 

\begin{figure}[t]
\centering
\includegraphics[width = 0.4\textwidth]{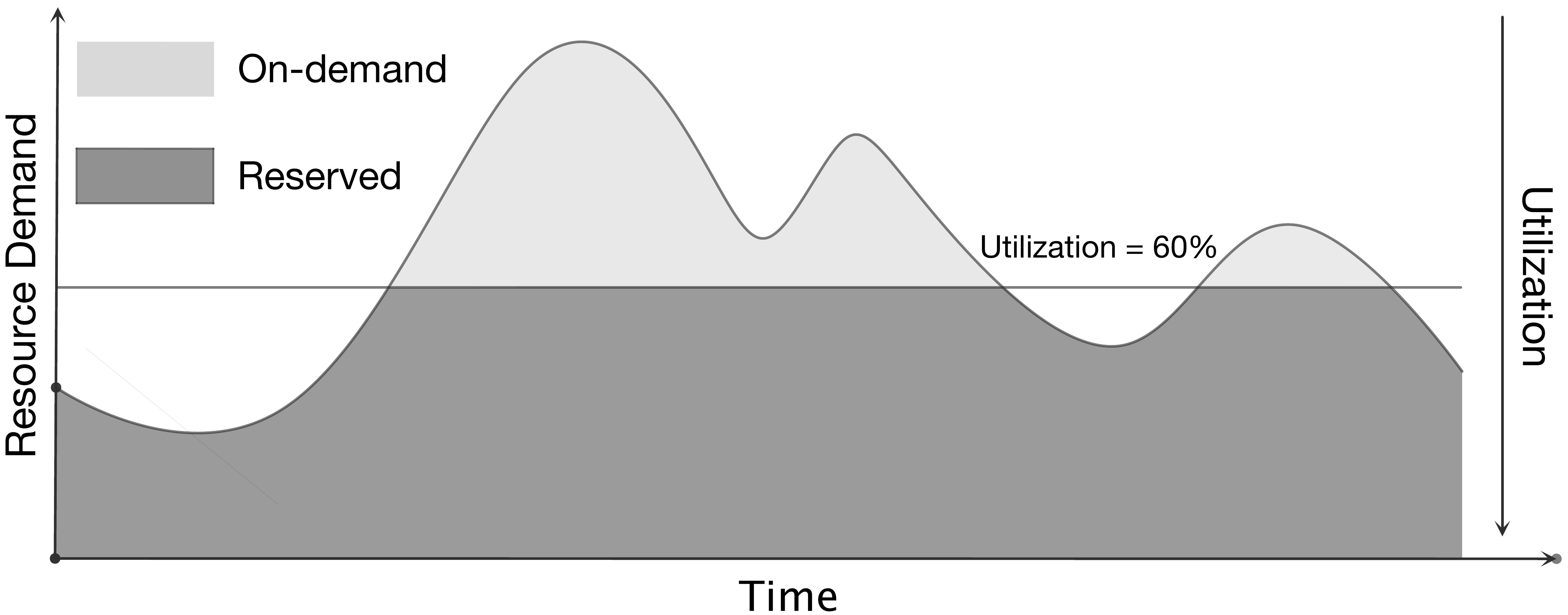}
\vspace{-0.3cm}
\caption{\emph{Illustration of the utilization of each unit of resource demand for normalizing the reserved option cost.}}
\vspace{-0.5cm}
\label{fig:example}
\end{figure}

\noindent {\bf Reserved}. In the offline case, when normalizing the cost of the reserved option, we assume a fractional resource, as mentioned above, that aggregates the resource demand of all jobs.  Figure~\ref{fig:example} shows an illustrative example of the aggregate resource demand over time for a batch trace. For each unit of stacked resource demand starting at $0$ on the y-axis, we compute the utilization of that unit of demand over both the 1- and 3-year offered reserved terms.  For the 1- and 3-year terms we assign the same normalized cost per unit of time to each unit of stacked resource demand starting at 0 based on its utilization across the term.  This has the effect, as in Figure~\ref{fig:example}, of drawing a line at each unit of demand on the y-axis, computing its utilization, and then normalizing its cost across the term length based on the utilization. 

Since the 1- and 3-year terms are 60\% and 40\% of the on-demand cost as shown in Table~\ref{table:options}, the resource utilization must be at least 60\% and 40\%, respectively, to have a normalized cost less than the on-demand cost.  The higher the utilization, the higher the normalized cost, such that 100\% utilization yields the entire discount. The 3-year term's cost is competitive with the transient option for long jobs if its utilization is high, e.g., 90-100\%, while the 1-year term's cost is competitive with spot blocks at high utilizations. The choice of selecting a 1- versus a 3-year term is a function of the expectations of each unit of demand's utilization over the term. As we show in \Section\ref{sec:evaluation}, since our 4-year batch trace has a mostly stable year-to-year demand, the 3-year term dominates in the offline case. 

\noindent {\bf Scheduled Reserved}.  Computing the normalized cost of the scheduled reserve option is similar to that of reserved, in that we normalize the cost relative to the utilization over the scheduled reserved's 1-year term.   However, unlike with the reserved option, which is essentially a single schedule, with scheduled reserved we must consider every possible daily, weekly, and monthly schedule.  For example, for a daily schedule, to meet the 1200 hour yearly minimum, we must reserve at least 4 consecutive hours per day starting at the beginning of an hour.  Thus, there are 21 possibly 4-hour schedules, 20 possible 5-hour schedules, 19 possible 6-hour schedules, etc., for a total of 210 possible daily schedules (note that mixing schedules of different lengths will further increase the number of possible daily schedules).   A weekly schedule is similar to a daily schedule, but only runs on certain days of the week, e.g., Monday and Friday, while a monthly schedule only runs on certain days of the month, e.g., the 1st and 15th.  There may be multiple non-overlapping schedules for each unit of demand over a 1-year term that yield a lower cost than either the on-demand or reserved price. 

To find these schedules, we observe that the problem reduces almost directly to the classic weighted job scheduling problem, which has an efficient and well-known dynamic programming solution that runs in $O(n\log{}n)$ time (and is often used in tutorials to teach dynamic programming\footnote{\url{https://www.geeksforgeeks.org/weighted-job-scheduling-log-n-time/}}). The weighted job scheduling problem takes as in put a set of $n$ jobs that each have a start and end time, as well as an associated value.  Given multiple, possibly overlapping jobs in time, the problem is to select the non-overlapping set of jobs, such that we maximize overall value. In our approach, each possible multi-hour scheduling interval is akin to a job, the normalized discount of that schedule is akin to the value, and the output is the cheapest set of schedules.  Thus, we can solve the problem directly, given the normalized cost for each schedule.    

For daily schedules, for each unit of resource demand, we simply compute its average utilization for each hour of each day over the year, which yields 24 values.  Given these 24 average utilization values, we can compute the average utilization for any multi-hour daily schedule interval.  This enables us to compute any daily schedule's normalized cost based on its utilization over the year and its discount.  Note that when computing the normalized cost, we compare it to the 1-year reserved cost for that unit of demand and discard the schedule interval if its cost is higher than the 1-year reserved cost.  For weekly schedules, we simply compute the hourly average utilization over the year for each day of the week, and compute the normalized cost for all possible schedules over the week (with different day combinations).  We apply the same approach for monthly schedules.   While weekly and monthly schedules admit many possible schedule combinations, the minimum length of a reservation (which is based on satisfying 1200 aggregate hours in a year) combined with the maximum number of days in a week and month make this number tractable.  The number of weekly schedules is 2025, and while the number of monthly schedules is $\sim$2B, many are discarded due to high prices and not used as input to the algorithm.

\noindent {\bf Selecting Purchasing Options.}  The policies above yield a normalized cost for each VM purchasing option for each unit of demand at each unit of time.  For the transient and spot block options, we sort their costs, such that the first unit of stacked resource demand at each time unit always has the lowest transient and spot block cost.  In our optimistic offline case, when selecting the purchasing option, we ignore that these costs map to individual jobs.   To compute our optimistic  offline optimal cost, we select the lowest cost for each unit of stacked resource demand, starting at $0$.  We first determine the lowest of the normalized transient, spot block, on-demand, and scheduled reserved options for each unit of resource demand for each unit of time before considering the reserved options.

For the reserved options, we first consider the 1-year option.  We compute the average cost of the lowest cost non-reserved options above for each unit of resource demand over each 1-year term.   That is, we determine for each unit of stacked resource demand, whether the 1-year reserved option or the average cost of the other non-reserved options yields the lowest cost.   Since we can purchase a reserved option at any time, we use a 1-year sliding window that performs this comparison over each 1-year interval in our data.  Assuming there is $>$3 years of data available, we next apply the same approach to compute the normalized 3-year reserved cost.  We compare this normalized 3-year cost with that of the lowest of the 1-year reserved option and non-reserved options, and take the lowest value.  We can apply a sliding window depending on data availability. Our approach above will yield the lowest cost purchasing option for each unit of stacked resource demand over time under our assumptions of perfect future knowledge and fractional supply and demand.  

\subsection{Practical Online Approach}
\label{sec:online}

Since our optimistic assumptions for the offline approach  are not practical, we adapt it to an online approach and evaluate it in \Section\ref{sec:evaluation} using 4 years of job submission data from a large-scale batch cluster.  Our practical online approach is essentially the same as our offline approach, but utilizes predictions of short- and long-term demand in the place of perfect knowledge, and does not assume a fractional supply and demand.  That is, each job requests a certain number of cores and memory, and (with the exception of the customized option) each VM type has a specific number of cores and memory.  Since our predictions are imperfect, our online approach is a heuristic.   However, even given perfect future knowledge of the workload, the problem is NP-hard, as removing the fractional assumption makes it strictly harder than the NP-hard bin packing problem. 

Our predictions of long-term demand are straightforward: we simply take prior job submission data and apply our offline approach from the previous section to estimate the amount of 1-year, 3-year, and scheduled reserved capacity to purchase. In \Section\ref{sec:evaluation}, we make these decisions based on the first year of job data, and evaluate our online approach over the next 3 years.   Since we do not have 3 years of prior data, we simply assume our training year will repeat to estimate the 3-year reserved capacity to purchase. The accuracy of such long-term predictions is a function of whether a workload changes significantly over time.  In many cases, such changes are often the result of exogeneous factors not present in historical data, such as new users gaining access. Thus, this (or any) large cluster's operators with knowledge of these exogeneous factors may be able to develop a better prediction model. Our goal here is not to develop the most accurate prediction model, rather it is to quantify the long-term cost benefits of mixing different contract types using reasonable predictions. 

The offline approach separately determines the amount of cores and memory to purchase under each of the reserved purchasing options.  In practice, we must map these cores and memory to specific types of VMs.  To do this, we simply purchase the largest VM types available that have a ratio of cores to memory that is closest to the offline ratio for each purchasing option.  Note that the per-resource on-demand cost of different VM types (with the exception of specialized types, such as GPUs or FPGAs, which we do not consider) is similar, so we do not consider cost in selecting VM types.  We are biased towards large VMs, since they are can run larger jobs.  

\begin{figure}[t]
\centering
\includegraphics[width = 0.35\textwidth]{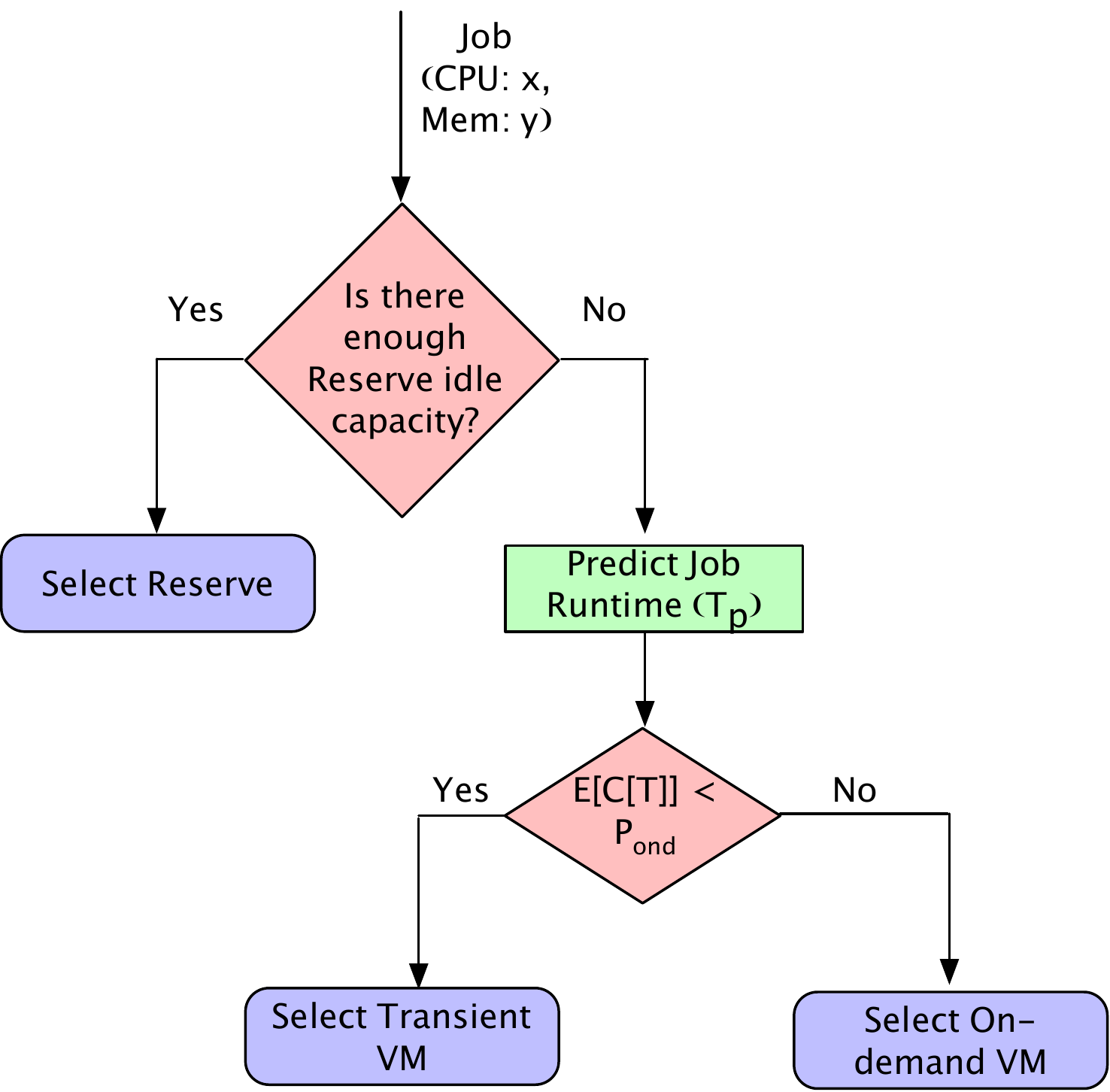}
\vspace{-0.5em}
\caption{\emph{Simple flow chart for selecting the VM purchasing option online when only reserved, transient, and on-demand are available, as with Microsoft}}
\vspace{-0.75cm}
\label{fig:flow-chart}
\end{figure}

After purchasing reserved capacity at the outset, as jobs arrive online, we schedule them on available reserved capacity based on their requested cores and memory.  If there are no available resources to execute a job, we dynamically acquire additional non-reserved resources to execute the job.  Since we are using the cloud, our batch system is not limited to a fixed-size cluster, and thus jobs never need wait in a queue for resources. When dynamically acquiring non-reserved resources, we must determine whether to purchase on-demand, transient, or spot block.\footnote{The sustained-use discount is always automatically applied to on-demand when used, so we need not consider it here.} Since the normalized cost of each is a function of the job's duration, we use training data to develop a simple model to predict job runtimes, which we discuss below.  

Given a prediction of the job's running time (as well as the transient revocation characteristics), we compute the normalized cost of each option and select the lowest cost.  We compute this normalized cost for every available VM type based on the job's requested cores and memory.   The lowest cost VM type is generally the smallest one that has the requisite cores and memory, although a discrete set of VM types results in wasted resources that increase the normalized cost.  The customized option eliminates this waste but at a 5\% increase in cost.  We also use the prediction to determine whether we can schedule a job to run on a scheduled reserved VM, such that it completes before the scheduled reservation expires.   Figure~\ref{fig:flow-chart} depicts a simple flow chart for the online case where only reserved, transient, and on-demand options are available (as with Microsoft).  Here, $E[C[T]]$ is the normalized cost to execute a job of length $T$ on a transient VM, while $P_{ond}$ is the expected cost to execute it on an on-demand VM. 

\noindent {\bf Job Runtime Predictions.} We develop a simple regression model based on a year of historical job submission data to predict job runtime.  As above, our goal is not to develop the most accurate job runtime prediction model, but to quantify the long-term cost benefits of mixing different contract types using reasonable predictions. Each job in our batch trace, described in \Section\ref{sec:implementation}, lists a user ID, job submission time, requested cores and memory, and maximum runtime limit.  The maximum runtime limit is supplied by the user and represents the maximum time the job can run before the system kills it---it is not a job runtime estimate.  We use these attributes as the input features to a regression model with the job runtime as the output variable. Once trained, the model supplies a job runtime prediction given a job's input features.  We assume this prediction is accurate when estimating the normalized cost of the on-demand, transient, and spot block options.


\section{Implementation}
\label{sec:implementation}
We implemented both the optimistic offline approach (\Section\ref{sec:offline}) and practical online approach (\Section\ref{sec:online}) in Python.  The offline implementation takes as input a trace of job submissions, and uses it to compute the mix of VM purchasing options that minimize the cost based on the assumptions in \Section\ref{sec:offline}. Each job entry includes its submission time, requested number of cores and memory, and running time.  The online implementation also takes as input a prior year's trace of job submissions, and uses it to determine the amount of 1-year, 3-year, and scheduled reserved capacity, assuming subsequent years will be similar to the prior year.  The online implementation also regresses on this data to build its job runtime prediction model.

We evaluate our approach in simulation using a 4-year trace of job submissions from a 14k batch cluster for a major state University system (serving multiple campuses).  Our simulation requires that jobs always receive the cores and memory they request, and have the same runtime as in the original trace. We analyze this trace in \Section\ref{sec:trace} to highlight important characteristics that impact the saving of different purchasing options.  In addition to the job submission time and requested cores and memory, each job entry also includes a user ID and maximum running time limit.  For job runtime predictions, we use ridge regression using these 4 input features and a job's actual running time as the output feature.  For the online approach, we use the first year of jobs (2015) for training, and then evaluate on the next 3 years (2016-2018) of jobs.  We evaluate the offline approach on the same 3 years.

\section{Evaluation}
\label{sec:evaluation}
%
%
%
%
%

Our evaluation examines the cost benefits of using a mix of VM purchasing options in both the offline and online cases compared to using a single purchasing option, either all on-demand or all reserved, for our batch trace.  We examine the cost benefits for the set of purchasing options offered by each cloud provider.  Specifically, Microsoft offers on-demand, 1- and 3-year reserved, and transient.  Google offers the same as Microsoft but also with a sustained-use discount and a customized option, while Amazon also offers the same as Microsoft but with scheduled reserved and spot block.  In addition, Google's variant of the transient option has a maximum lifetime of 24 hours, while Amazon and Microsoft's variant has no maximum lifetime.  Thus, Amazon and Microsoft offer a transient option with longer potential times-to-revocation.  For our experiments, we assume an average time-to-revocation of 12 hours for Google (with a maximum of 24 hours) uniformly distributed, while for Microsoft and Amazon, we assume an average of 48 hours (based on an analysis in prior research~\cite{flint}).   Since the revocation characteristics of the transient option are not well-known, we also evaluate the mix without considering the transient option.
 
Since our focus is on the benefits of different purchasing options, we use the same standard set of VM types and prices across all providers.  As a result, our evaluation does not reflect the absolute cost difference between providers, but the relative benefits of each provider's set of purchasing options. We consider standard VM types with 1, 2, 4, 8, 16, 32, and 64 cores with 4, 8, 16, 32, 64, 128, and 256 GB memory, respectively. We assume the cost of a 1 core, 4GB VM is \$0.0481 per hour, which is equivalent to an {\tt m5.large} VM in Amazon with larger capacity VMs priced as a simple scalar multiple.  Google and Microsoft's quoted prices for a similar size VM are roughly the same. Below, we highlight salient characteristics of the batch trace that impact our results before evaluating the benefits of mixing VM purchasing options. 

\begin{figure}[t]
\centering
\includegraphics[width=0.45\textwidth]{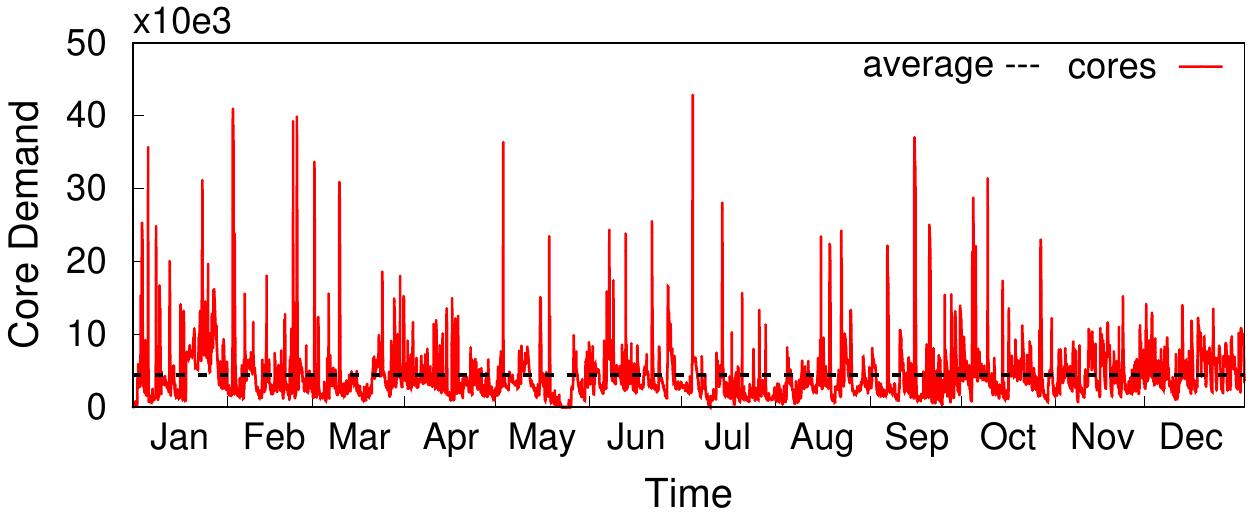}
\vspace{-0.5cm}
\caption{\emph{The hourly core demand in 2018 with an average of 4380 cores.}}
\vspace{-0.5cm}
\label{fig:hourly_usage}
\end{figure}

\subsection{Batch Trace Characteristics}  
\label{sec:trace}

Figure~\ref{fig:hourly_usage} shows the hourly core demand on average for our batch cluster over the year 2018.  While our cluster has 14k cores, the core demand peaks at nearly 43k cores, indicating that jobs may periodically experience long waiting times.  In the cloud, these waiting times are not necessary as there is no resource constraint. As might be expected, the average core utilization over the year is much less than the peak, at only 4380 cores, resulting in a 31\% average utilization.  The high peak-to-average demand ratio and the low average utilization make our batch workload well-suited for the cloud, which has the potential to reduce both job waiting times (by acquiring more resources when demand is high) and cost (by releasing non-reserved resources when demand is low). 

\begin{figure*}[t]
\centering
\begin{tabular}{ccc}
\includegraphics[width=0.4\textwidth]{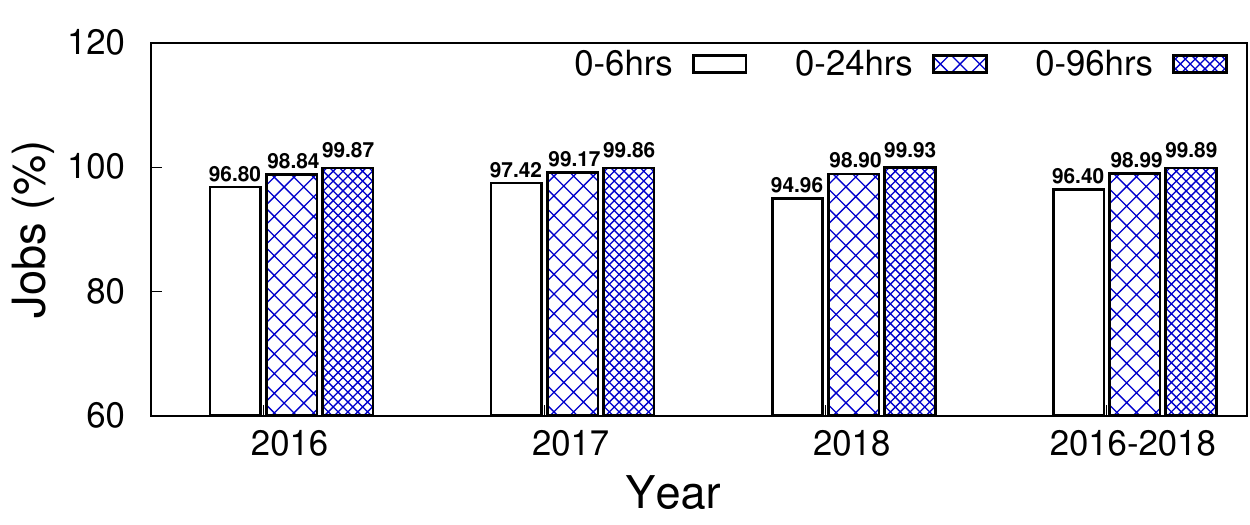}&
\includegraphics[width=0.4\textwidth]{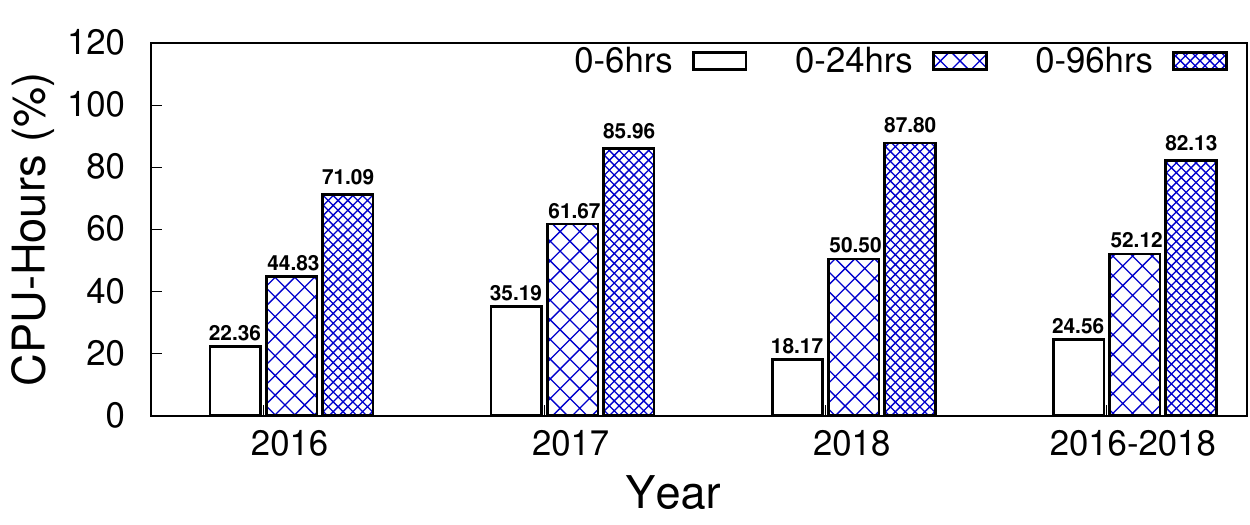}\\
(a)&
(b)\\
\end{tabular}
\vspace{-0.2cm}
\caption{Job runtime (a) and CPU-hours (b) for different length jobs each year in our batch trace.}
\vspace{-0.5cm}
\label{fig:job_categories_cdf}
\end{figure*}


Figure~\ref{fig:job_categories_cdf} shows a breakdown of jobs based on their running time (a) and core-hours (b) for each year (2016-2018) of our trace.   We break jobs into 3 categories based on their runtime: i) 0-6 hours, ii) 0-24 hours, and iii) 0-96 hours.  We select these runtime categories because they roughly correspond to the running times necessary to use spot blocks, Google transient preemptible VMs, and Amazon transient spot VMs.  That is, spot blocks only permit blocks from 1-6 hours, preemptible VMs have a maximum lifetime of 24 hours, and prior work suggests that Amazon spot VMs have an average lifetime of $\sim$48 hours but can run much longer without a revocation~\cite{flint}.

Figure~\ref{fig:job_categories_cdf}(a) shows that a high percentage ($>$96\%) of the jobs are less than 6 hours in length, and only a small percentage are longer.  The number of jobs that run longer than 96 hours is quite small at $0.11$\%.  However, Figure~\ref{fig:job_categories_cdf}(b) shows the CPU-hours the jobs in these categories consumes.  Importantly, even though $>$96\% of jobs are small, they only consume less than 25\% of the CPU-hours on average.  In contrast, the small number ($\sim$1\%) of 0-24 hour jobs comprise 52\% of the CPU-hours on average, while 0-96 hour jobs comprise 82\%. Thus, even though jobs longer than 96 hours comprise only $0.11$\% of the workload, they use 18\% of the CPU-hours over the 3-year period.  As we discuss below, these job characteristics impact the the cost savings of different purchasing options.

\subsection{Mixing VM Purchasing Options}

\begin{figure*}
\centering
\begin{tabular}{cc}
\includegraphics[width=0.4\textwidth]{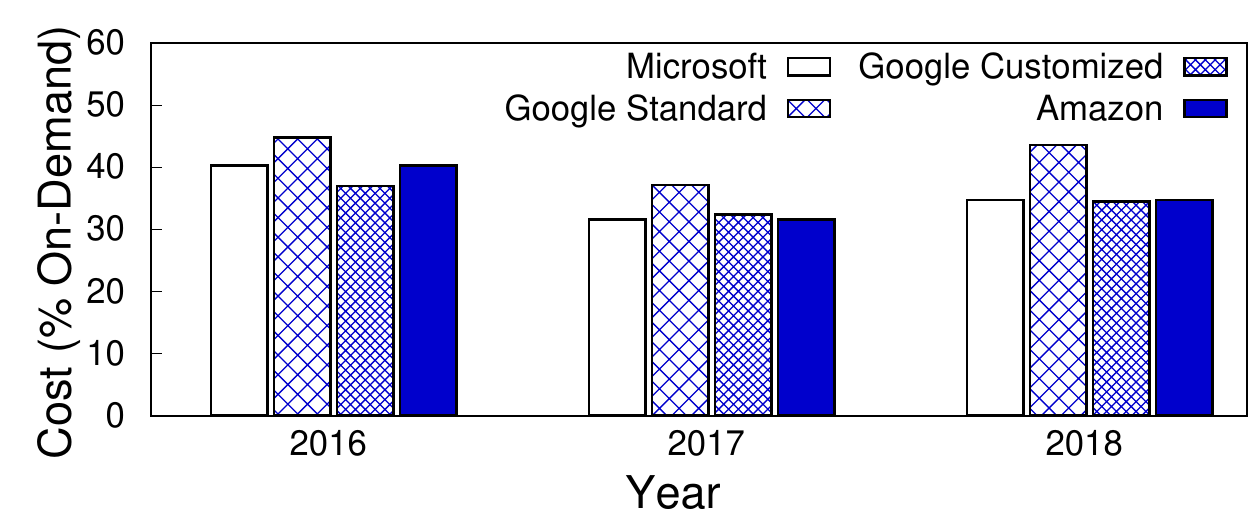} &
\includegraphics[width=0.4\textwidth]{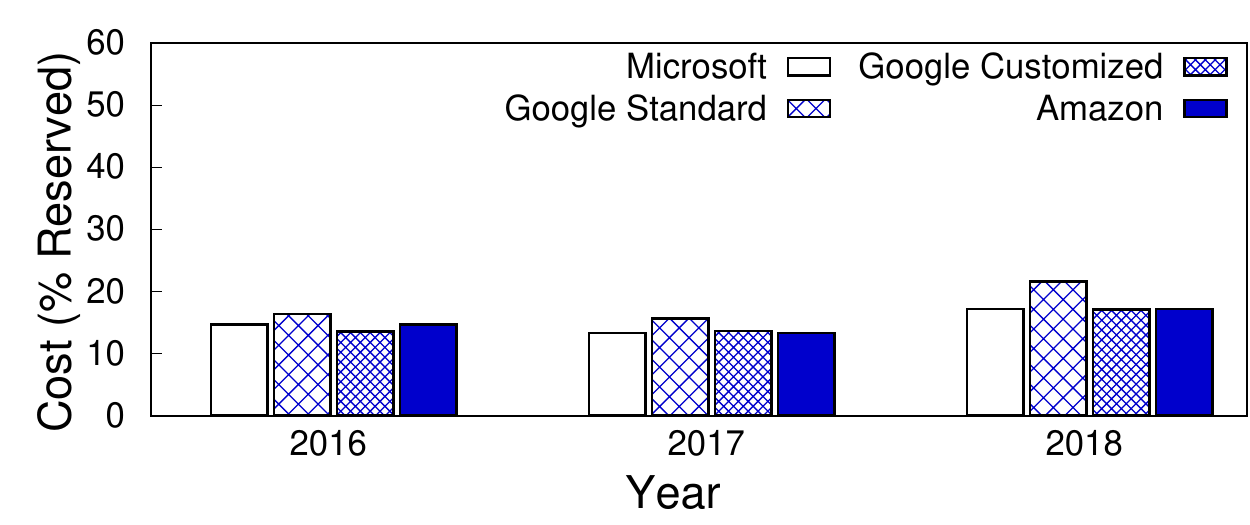} \\
 (a)&
 (b)\\
 \end{tabular}
 \caption{Cost for executing our batch trace using all purchasing options from the different cloud providers in the optimistic offline case as a percentage of using on-demand only (a) and reserved only (b).}
\vspace{-0.5cm}
\label{fig:offline_analysis}
\end{figure*}

\begin{figure}[h]
\centering
\includegraphics[width = 0.45\textwidth]{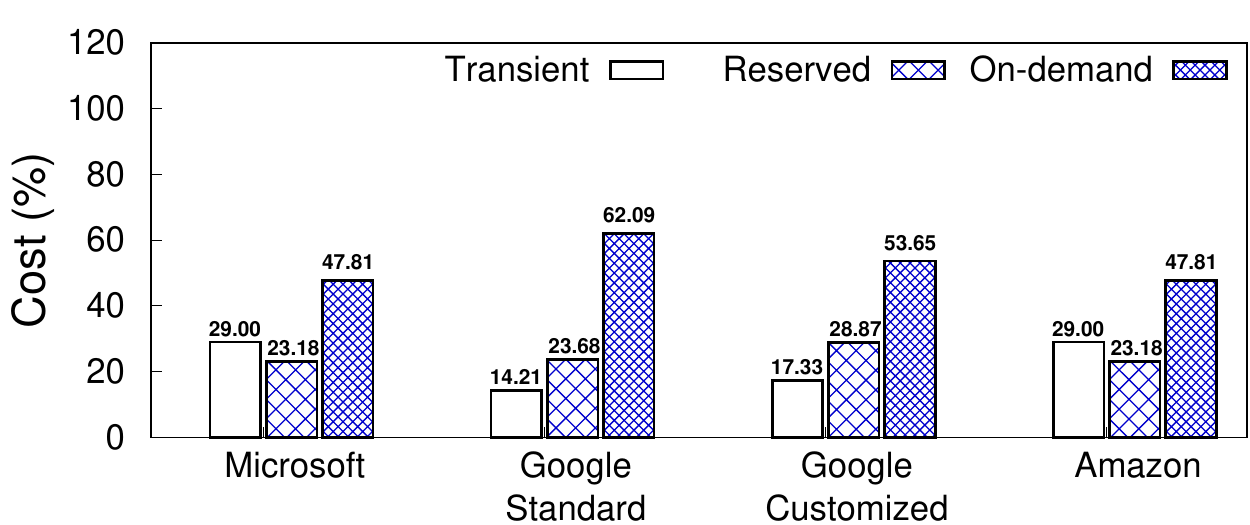}
\vspace{-0.5em}
\caption{Mix of VM purchasing options used over 2016-2018 in the offline case (with the transient option).}
\vspace{-0.5cm}
\label{fig:mix-fraction-offline}
\end{figure}

Below, we evaluate the cost savings from mixing all the VM purchasing options in both the online and offline case. 

\noindent {\bf Optimistic Offline Approach}. Figure~\ref{fig:offline_analysis} shows the cost of mixing VM purchasing options using our optimistic offline approach from 2016-2018 for the sets of VM purchasing options offered by the different cloud providers.  Figure~\ref{fig:offline_analysis}(a) plots the cost relative to using only on-demand VMs, while (b) plots the cost relative to reserving enough VM capacity to satisfy the peak demand with no job waiting time.  Figure~\ref{fig:mix-fraction-offline} shows the average percentage mix of each purchasing option over the 3-year period.  We see that the cost relative to on-demand is 35\% for Amazon and Microsoft, but only 41\% for Google-Standard, which includes the sustained-use discount but does not permit the customized option.  The savings for Google-Customized drops to 33.62\% of the on-demand cost. 

The Amazon and Microsoft cases are the same because Amazon's additional options---spot block and reserved---are never used in the offline case, and we treat the transient case the same for both Microsoft and Amazon.  Spot blocks never offer a cost benefit for short 1-6 hour jobs over transient, although they may be used in practice based on a job's requirements, i.e., if it cannot risk a revocation. Scheduled reserved is never selected, largely because its discount is too low (5-10\%). In particular, there is no unit of stacked demand with a daily, weekly, or monthly period where the utilization within the period is $>$90-95\%, but where the utilization overall for that unit of demand is $<$60\%, which would result in that unit of demand being satisfied by a reserved VM.  Overall, transient VMs dominate the offline mix with $\sim$29\% because they yield a much lower expected cost, even when considering the likelihood of revocation and when using our simple policy of always running a revoked job on on-demand.   However, despite their low cost, non-transient purchasing options still account for $\sim$71\% of the workload. 

Google's set of purchasing options is cheaper due to both their sustained-use discount and customized option.  Adding the sustained-use discount in Google-Standard results in a cost 41\% the on-demand cost, while also adding the customized option reduces this to 33.62\%, despite the 5\% increase in price. Nearly every job makes use of the customized option, as most are not within 5\% of the size of a standard VM.  The customized option enables us to match the cores requested by a job to the closest multiple of 2, rather than the closest power of 2 (as with Amazon and Microsoft).  In addition, Google enables users to allocate up to 6.5GB per core, while the standard offerings from Amazon and Microsoft associate 4GB with each core. Since a large number of jobs in our workload have $>$4GB memory per core, this increases the relative benefit of the customized offering over the standard VM types.  Finally, Figure~\ref{fig:offline_analysis}(b) shows that using only the 1-year reserved option is much more expensive than only using the on-demand option despite its 40\% discount.  As expected, this is due to our batch trace's high peak-to-average ratio in demand. 

\begin{figure*}
\centering
\begin{tabular}{ccc}
\includegraphics[width=0.4\textwidth]{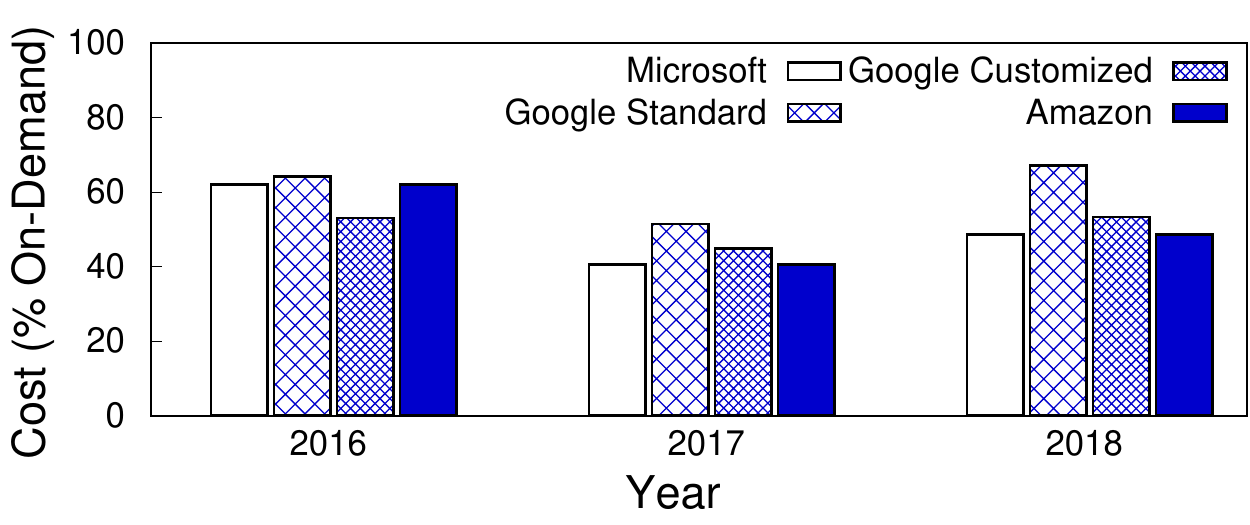} &
\includegraphics[width=0.4\textwidth]{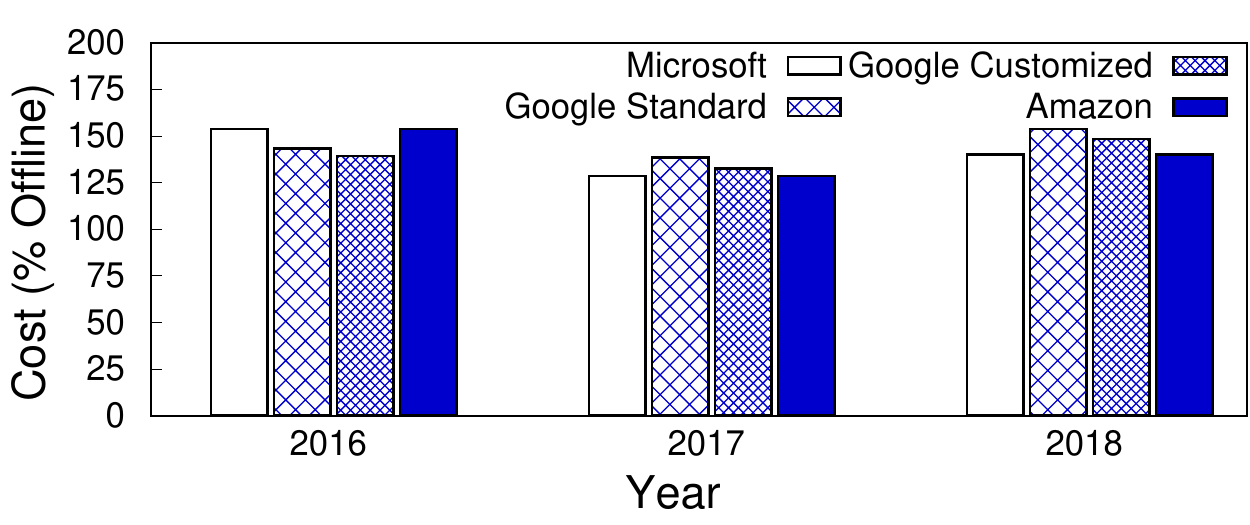} \\
 (a)&
 (b)\\
 \end{tabular}
 \vspace{-0.2cm}
 \caption{Cost for executing our batch trace using all purchasing options from the different cloud providers in the online case as a percentage of using on-demand only (a) and reserved only (b).}
\vspace{-0.7cm}
\label{fig:online_analysis}
\end{figure*}

\begin{figure}[h]
\centering
\includegraphics[width = 0.45\textwidth]{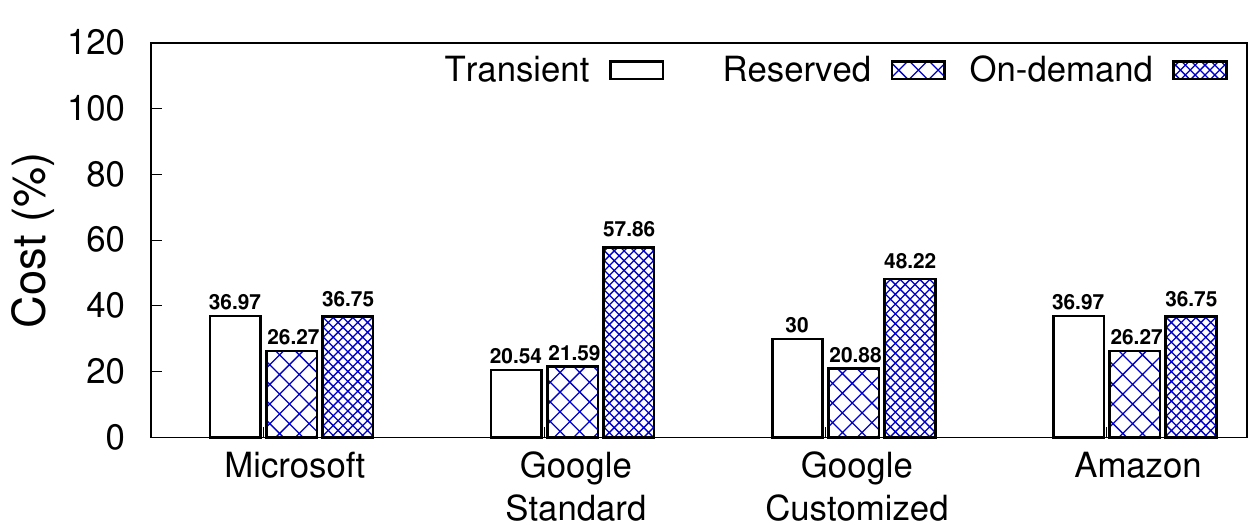}
\vspace{-0.5em}
\caption{Mix of VM purchasing options used over 2016-2018 in the online case (with the transient option).}
\vspace{-0.5cm}
\label{fig:mix-fraction-online}
\end{figure}

\noindent {\bf Practical Online Approach}. Figure~\ref{fig:online_analysis} shows the cost of our online approach both as a percentage of the cost of only using the on-demand option (a), and as a percentage of the optimistic offline cost (b).  Figure~\ref{fig:mix-fraction-online} shows the average percentage of each VM purchasing option over the 3-year period.  The mix of purchasing options is similar to the offline approach above, although the percentage of transient VMs used decreases due to inaccurate job runtime predictions, as we discuss below.  As above, the online approach never selects the spot block option or uses scheduled reserved.  As a result, both Amazon's and Microsoft's results are the same in the online case as well.  Also as before, Google-Standard and Google-Customized results in slightly lower relative costs in (a).   In the online case, however, Google's set of purchasing options do not yield quite as much relative benefit as in the offline approach.  This occurs because the offline case knows all job runtimes, while the online case has to predict them.  Since Google has a lower maximum lifetime (24 hours) for its transient option, incorrect job runtimes (which, in our case, tend to predict jobs to be shorter than they actually are) have a greater impact on Google in the online case.  That is, Google has more jobs assigned to transient VMs that get revoked due to incorrect predictions, which increases their overall cost.  These incorrect predictions reduce Google's cost advantage to 39\% (to 69\% the on-demand cost verses 50\% for Amazon and Microsoft) across the 3-year period. 

Figure~\ref{fig:online_analysis}(b) compares the online approach for each set of purchasing options with the respective offline approach.  As mentioned above, Google's online cost is the highest relative to its offline cost due to incorrect job runtime predictions combined with its short maximum lifetime for transient VMs.  The online approach results in 35\% greater cost compared to their respective optimistic offline approach for Amazon and Microsoft, while it results in 55\% greater cost for Google. 

\subsection{Removing Transient VMs}


\begin{figure}[h]
\centering
\includegraphics[width = 0.45\textwidth]{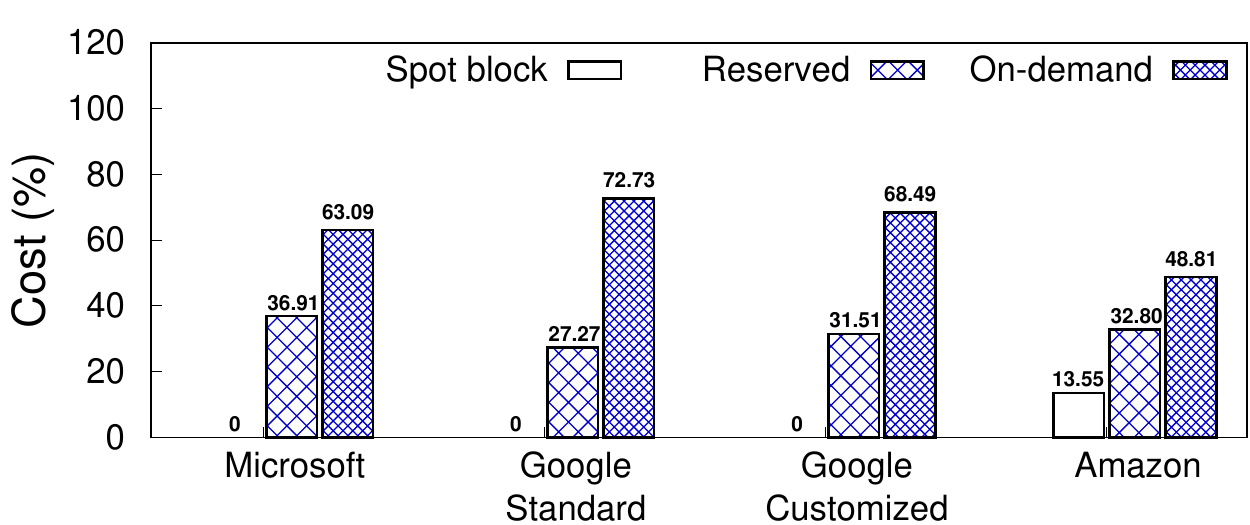}
\vspace{-0.5em}
\caption{Mix of VM purchasing options used over 2016-2018 in the offline case without the transient option.}
\vspace{-0.7cm}
\label{fig:mix-fraction-offline-notransient}
\end{figure}

We repeat the evaluation above, but in this case without the transient option.  We do this for two reasons.  First, transient characteristics are not publicly known, and have a significant impact on our results.  Second, not all jobs can run on transient VMs due to having strict deadlines or not being able to gracefully handle revocations. 

\noindent {\bf Optimistic Offline Approach}.  Figure~\ref{fig:mix-fraction-offline-notransient} shows the cost of mixing VM purchasing options without the transient option in the optimistic offline case. The figure shows that the overall cost increased relative to with the transient option because transient VMs were by far the cheapest option. However, this does not occur to a significant degree, primarily because not a large fraction of the CPU-hours come from jobs that are less than 6 hours in length (see Figure~\ref{fig:job_categories_cdf}(b)), which mitigates the impact of spot block.  However, the availability of spot block for Amazon with no low cost replacement for transient at either Google or Microsoft results in Amazon having the lowest overall cost.  Their cost savings exceed Google even when using the sustained-use discount and customized option. 

\begin{figure}[h]
\centering
\includegraphics[width = 0.45\textwidth]{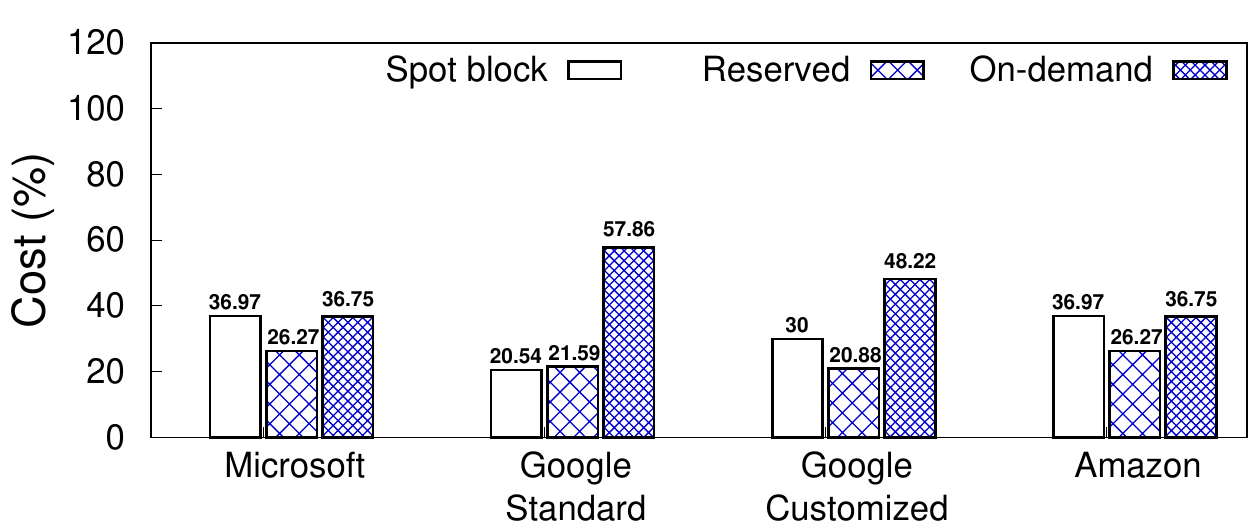}
\vspace{-0.5em}
\caption{Mix of VM purchasing options used over 2016-2018 in the online case without the transient option.}
\label{fig:mix-fraction-online-notransient}
\end{figure}

\noindent {\bf Practical Online Approach}.  Figure~\ref{fig:mix-fraction-online-notransient} then shows the same results when using our online approach as a percentage of the on-demand cost.  The benefits of spot block decrease when using the online approach, since the predictions of job running times are not accurate.  Due to the reduced benefit of spot block, Google's sustained-use discount and customized option enable it to achieve the lowest cost.   Figure~\ref{fig:mix-fraction-online-notransient} shows the mix of purchasing options, which has a slightly reduced percentage of spot block VMs compared to Figure~\ref{fig:mix-fraction-offline-notransient}.


\section{Related Work}
\label{sec:related}
There has been significant prior work in optimizing particular workloads and applications for using different VM purchasing options.  However, most of this work focuses on optimizing a single type of purchasing option and comparing it with using on-demand, rather than looking at all of them in combination.  Perhaps most relevant to our work is HCloud~\cite{hcloud}, which focuses on combining the reserved and on-demand purchasing options. However, HCloud anecdotally shows that reserved VMs tend to have less performance interference than on-demand VMs of an equivalent type, although cloud SLOs do not specify a difference between purchasing options of equivalent types of VMs.   HCloud primarily focuses on determining how to map jobs to on-demand and reserved VMs based on their sensitivity to performance interference.  Thus, HCloud targets small-scale workloads and does not evaluate their approach using a real large-scale workload over a multi-year period, nor optimize for other purchasing options.   

There is a large body of work on optimizing for other purchasing options, including transient VMs~\cite{tr-spark,flint,exosphere,spoton,pado}, the sustained-use discount~\cite{virtual-cloud}, and burstable VMs (which we consider to be a different VM type)~\cite{burstable}.  Thus far, we know of no work that examines the benefits of spot block, scheduled reserved, or customized VMs.  There is also a large body of related work on selecting the appropriate VM type for a particular application or workload~\cite{cherrypick,bestvm2,ernest,bestvm}.  Our work differs from this work in that we focus on selecting the best purchasing option given accurate knowledge of a job's resource requirements.   Of course, user resource requests for jobs may not be the most efficient.  Thus, our work could incorporate prior work on automatically determining the resources needed for a job to further optimize cost.

\section{Conclusion}
\label{sec:conclusion}
Cloud platforms offer the same VMs under a variety of purchasing options  that specify different costs and time commitments, such as on-demand, reserved, sustained-use, scheduled reserve, spot/preemptible, and spot block.  Choosing from among these options can be challenging.  To address this problem, in this paper, we  design policies to optimize long-term cloud costs by selecting a mix of VM purchasing options based on short- and long-term expectations of workload utilization. We evaluate our policies on a batch job trace spanning 4 years from a large shared cluster for a major state University system that includes 14k cores and 60 million job submissions, and show how these jobs could be cost-effectively executed in the cloud using our approach. Our results show that our policies incur a cost within 41\% of an optimistic offline optimal approach, are 50\% less than solely using on-demand VMs, and 79\% less than using reserved VMs. 
\\
\noindent {\bf Acknowledgements.} This work was funded by NSF grants \#CNS-1802523 and \#CNS-1908536.


\bibliographystyle{abbrv}
\bibliography{paper}

\end{document}